# Scalar, tensor and vector polarizability of Tm atoms in 532 nm dipole trap

V. V. Tsyganok[1, 2], D. A. Pershin[1, 2], E. T. Davletov[1, 2], V. A. Khlebnikov[1]
and A. V. Akimov[1,3,4]

[1]Russian Quantum Center, Business Center "Ural", 100A Novaya Str., Skolkovo, Moscow, 143025, Russia
[2]Moscow Institute of Physics and Technology, Institutskii per. 9, Dolgoprudny, Moscow Region 141701, Russia
[3]PN Lebedev Institute RAS, Leninsky Prospekt 53, Moscow, 119991, Russia
[4]Texas A&M University, 4242 TAMU, College Station, Texas, 77843, USA,
email: akimov@physics.tamu.edu

## I. ABSTRACT

Dipolar atoms have unique properties, making them interesting for laser cooling and quantum simulations. But, due to relatively large orbital momentum in the ground state these atoms may have large dynamic tensor and vector polarizabilities in the ground state. This enables the formation of spin-dependent optical traps. In this paper real part of tensor and vector dynamic polarizability was experimentally measured and compared to theoretical simulation. For an optical dipole trap operating around 532.07 nm tensor, polarizability was found to be $-145 \pm 53$ a.u. and vector was $680 \pm 240$ a.u. The measurements were compared with simulations, which were done based on the known set of levels from a thulium atom. The simulations are in good agreement with experimental results. In addition, losses of atoms from the dipole trap were measured for different trap configurations and compared to the calculated imaginary part of vector and tensor polarizabilities.

## II. INTRODUCTION

Ultracold atoms have high potential in the field of quantum simulations [1–3]. One of the key advantages of cold atomic ensembles is a large degree of control over interatomic interactions as well as the internal states of an atom [4]. Among other elements, the rare earths hold a special place on the periodic table as they have incomplete electronic f-shells and therefore also have high orbital and magnetic moments in the ground state. This affects many properties of rare-earth elements, including large number of Feshbach resonances in low fields [5,6] and strong dipole-dipole interactions [7]. Another important degree of control enabled with rare-earth elements is their highly anisotropic polarizability in a wide range of light spectrum, which is already well manifested even for atomic ground state.

Dynamic polarizability is an important property of an atom, to a high degree determining the interaction of an atom with a non-resonant light field. Alkali atoms are known to have mostly scalar polarizability in the ground state due to a s-type electronic shell in ground state. To the contrary, rare-earth elements have a non-zero orbital momentum in the ground state. This gives rise to considerable contribution of the tensor and vector polarizabilities - even in the ground state. In particular, polarizabilities for the erbium atom were recently calculated [8] and measured experimentally [9].

In this paper we experimentally study the dynamic polarizabilities of cold thulium atoms in a 532 nm dipole trap [10]. By manipulating the orientation of atomic ensemble polarization and polarization of the light field, we were able to extract tensor and vector components of the dynamic polarizability and compare them with simulations based on known transitions in thulium atoms [11]. We demonstrate that around this wavelength contributions of tensor and vector polarizability is quite significant, thus allowing formation of well-separated spin-depended traps. Besides this, we also specifically analyzed losses from the atomic trap depending on the mutual orientation of atomic polarization and light polarization and compared it with our simulations. We found that losses do not follow the behavior of the real element of polarizability, thus we excluded the simplest radiative loss mechanism for our trap.

## III. SIMULATIONS

In the presence of a non-resonant light field of frequency $\omega$, atomic energy levels undergo shift leading to trapping potential $U(\omega)$. It could be expressed as the sum of the scalar $U_s$, vector $U_\upsilon$, and tensor $U_t$ parts [12] as following [9]:

$$U(\omega) = -\frac{1}{2\varepsilon_0 c} I(r) \mathrm{Re}[\alpha_{tot}] = U_s + U_\upsilon + U_t$$

$$U_s = -\frac{1}{2\varepsilon_0 c} I(r) \mathrm{Re}[\alpha_s(\omega)]$$

$$U_\upsilon = -\frac{1}{2\varepsilon_0 c} I(r) \varepsilon \cos\theta_k \frac{m_J}{2J} \mathrm{Re}[\alpha_\upsilon(\omega)]$$

$$U_t = -\frac{1}{2\varepsilon_0 c} I(r) \frac{3m_J^2 - J(J+1)}{J(2J-1)} \cdot \frac{3\cos^2\theta_p - 1}{2} \mathrm{Re}[\alpha_t(\omega)]$$

(1)

Where $\varepsilon_0$ is the vacuum permittivity; $c$ is the speed of light; $I(r)$ is the laser intensity profile; $\varepsilon = |\vec{u}^* \times \vec{u}|$ is the ellipticity parameter with $\vec{u}$ as the normalized Jones vector; $\theta_p = \angle(\vec{E}, \vec{B})$ and $\theta_k = \angle(\vec{k}, \vec{B})$ (see Figure 2A); $m_J$ is the angular-momentum projection quantum number; $J$ is the total angular momentum of electrons; $\alpha_{tot}$ is the total atomic polarizability; $\alpha_s(\omega)$, $\alpha_\upsilon(\omega)$, $\alpha_t(\omega)$ are scalar, vector, and tensor dynamic dipole polarizabilities respectively.

Imaginary parts of polarization values set photon scattering rates given by similar expression:

$$\Gamma(\omega) = \frac{1}{\hbar\varepsilon_0 c} I(r) \left[ \mathrm{Im}[\alpha_s(\omega)] + \varepsilon\cos\theta_k \frac{m_J}{2J} \mathrm{Im}[\alpha_\upsilon(\omega)] + \right.$$
$$\left. + \frac{3m_J^2 - J(J+1)}{J(2J-1)} \times \frac{3\cos^2\theta_p - 1}{2} \mathrm{Im}[\alpha_t(\omega)] \right]$$

(2)

To calculate all parts of dynamic dipole polarizability, we follow the sum-over-state approach (see APPENDIX A). Energy levels and corresponding natural linewidths for dipole-allowed transitions were taken from NIST database [11].

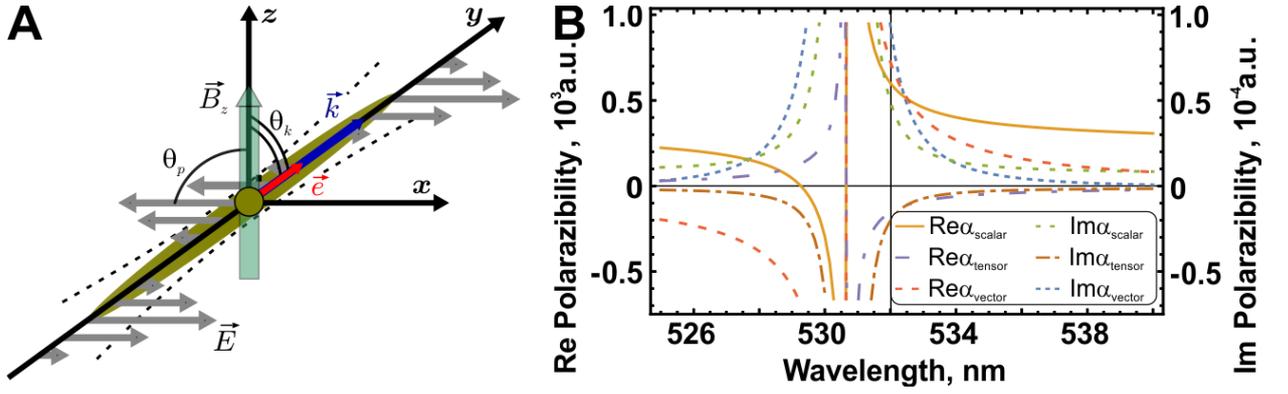

*Figure 1 A – Idea of the Experiment: polarized atomic cloud (orientated always along magnetic field) in a single-beam ODT with elliptical polarization that is propagate along y-axis ($\vec{e}$ vector). B – Real and imaginary parts of ground state polarizabilities in the region of 532.07 nm laser light. The black line is the wavelength of our ODT.*

Figure 1B depicts calculated values of the thulium atom's real and imaginary parts of ground state polarizabilities in the region of 532 nm. The experimentally measured frequency of the optical dipole trap (ODT) light was found to be 532.07 nm (using Wavelength Meter WS-7, calibrated by the thulium transition line). For this wavelength the simulation gives values of 583 a.u., 684 a.u., and -140 a.u. for real parts of scalar, vector, and tensor polarizabilities respectively; imaginary parts are $446 \cdot 10^{-7}$ a.u., $836 \cdot 10^{-7}$ a.u., $183 \cdot 10^{-7}$ a.u. resectively. These quantities are strongly affected by the near-lying 530.7 nm optical transition with a level's width of 330 kHz. Thus, it was found that vector and tensor part are almost entirely formed by this transition, while for scalar polarizability it provides about half of the value.

## IV. EXPERIMENT

To measure the polarizability of thulium at a wavelength of 532.07 nm, an atomic cloud of $^{169}$Tm was initially cooled down with a magneto-optical trap (MOT) [10]. In this type of MOT, the atomic cloud is spin-polarized to the lowest ground-state Zeeman sublevel $(J = 7/2, F = 4, m_F = -4)$ with a population of $m_F = -4$ (higher then 97%) [13]. Then atoms were transferred into a single-beam optical dipole trap operating at 532.07 nm [13]. After 300 ms of holding time in the optical dipole trap (ODT), about $1.5 \times 10^6$ atoms with a temperature of around 18 µK were typically achieved with vertical orientations of the magnetic field (along $z$ axis on Figure 2A) and linear horizontal polarization of the ODT beam $\theta_k \approx \theta_p \approx 90°$. To detect the atomic cloud, the absorption imaging technique was used [14].

In some experiments, $\theta_p$ was varied by changing currents in the magnetic field coils, thus turning the direction of the magnetic field. This rotation was performed after the light had been switched off. We were able to control the magnetic field values with 50 mG precision accuracy [13]. For angles different from $\theta_k \approx \theta_p \approx 90°$, the depth of the dipole trap changes a lot due to presence of tensor and vector polarizabilities (see Figure 2A and (1)). Thus, after such a rotation, the number of atoms as well as the temperature of the atomic cloud in the ODT varies greatly depending on the rotation angle. However, even in the worst condition for the ODT ($\theta_k = 90°, \theta_p = 0°$), we had enough atoms to perform an experiment. To understand the depth of the ODT $U$, the standard technique of trap-frequency measurements was used [9]. The total polarizability, $\alpha_{tot}$, then could be found using equation (1). For the Gaussian beam which propagates along the y-axis the intensity profile is:

$$I(x,z) = I_0 \exp\left[-\frac{2x^2}{w_x^2} - \frac{2z^2}{w_z^2}\right], \qquad (3)$$

where $w_x$ and $w_z$ are beam waists, and $I_0 = 2P/\pi w_x w_z$ where $P$ is beam power. The trap frequencies in harmonic approximation $\nu_i$ could be then calculated as [15]:

$$\begin{aligned}\nu_i &= \frac{1}{2\pi}\sqrt{\frac{-4U_0}{m_{th} w_i^2}}, \quad i \in \{x,z\} \\ \nu_y &= \frac{1}{2\pi}\sqrt{\frac{-2U_0}{m_{th} w_y^2}}\end{aligned}, \qquad (4)$$

where $m_{th}$ is the atomic mass, $w_y$ is Rayleigh length, and $U_0 = -\alpha_{tot} I_0 / 2\varepsilon_0 c$. Thus, using (4) one can find expression for total polarizability via parameters, measured in the experiment:

$$\alpha_{tot} = \pi^3 \varepsilon_0 c \frac{\nu_x^2 w_x^3 w_z m_{th}}{P} = \pi^3 \varepsilon_0 c \frac{\nu_z^2 w_z^3 w_x m_{th}}{P}. \qquad (5)$$

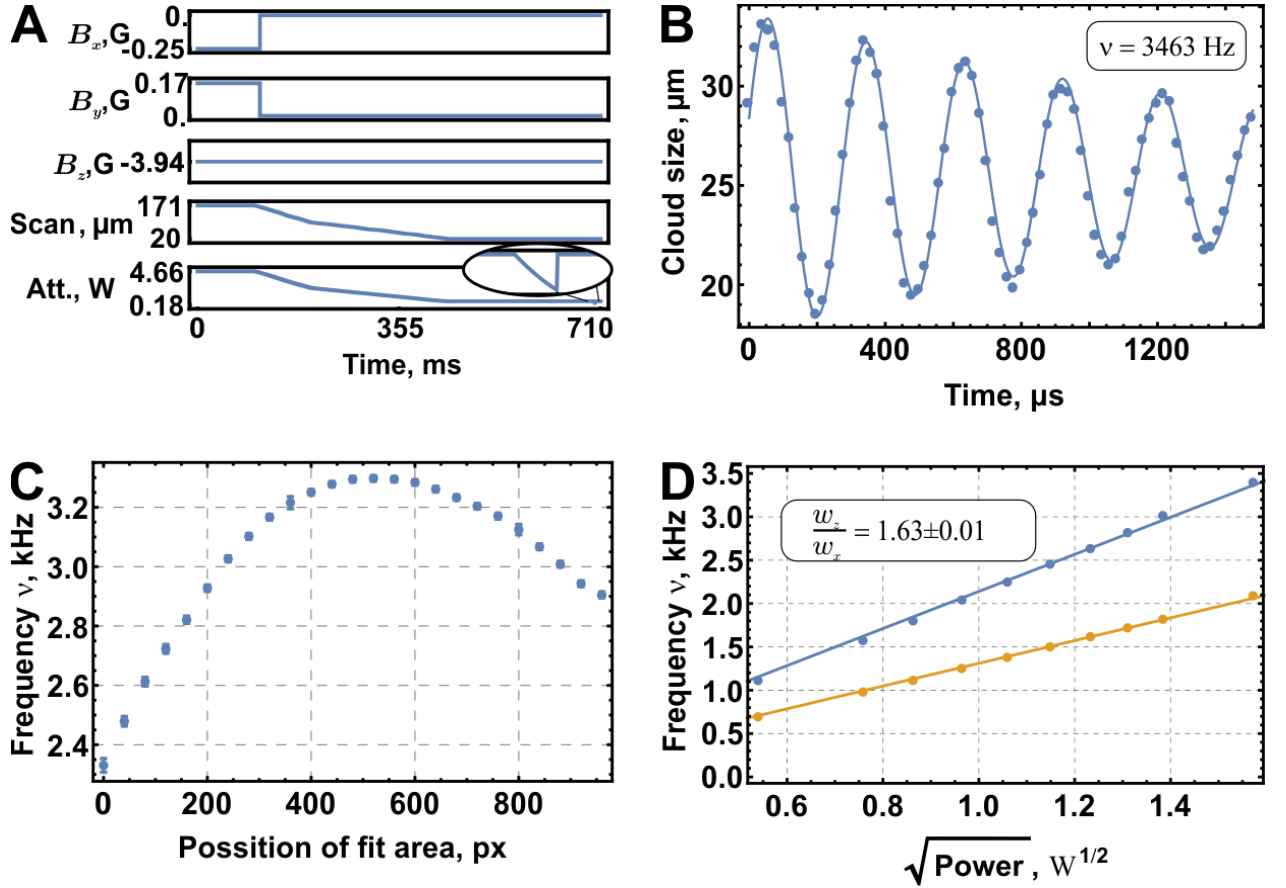

Figure 2 A – Scheme of pulses used in the experiment with tensor polarization. B – Typical fit of atomic cloud size oscillations versus time. C – Dependence of fitted frequency vs position of fit. D – Frequency versus square root of ODT-beam power (without sweeping) for both radial x- and z-axis.

## A. Frequency measurements

To measure the ODT frequency, an atomic cloud was kept in the ODT for 300 ms after ODT sweeping was tuned off [13]. After this time, evaporative cooling mostly stops and atoms can be considered thermalized. At this point, the atoms occupied the central part of the trap. The power of the beam was then decreased by 4 times over 2 ms, followed by a sharp increase to the required value of power; thereby, causing oscillations of atomic cloud size and position (Figure 2B).

Since the optical dipole trap was formed by a single Gaussian beam, the frequency of oscillations depends on the fitted region along the beam. Therefore, we divided the experimental images into 25 parts of 40 pixels each. For each region the frequency was fitted (see Figure 2B) and plotted versus position along the beam (Figure 2C). The maximum frequency found this way was used for calculating the trap frequency, which came out to be just ½ of the cloud size oscillation frequency.

## B. Beam waist

To measure the ODT waists, we used a CMOS Thorlabs DCU223M-GL camera by placing it into the laser beam reflected with an additional mirror in front of the vacuum chamber. The problem was that the size of the laser spot occupied a small number of camera pixels, resulting in a large inaccuracy of ODT waists. To overcome this, we performed our measurement with a sweeping trap [13], which has an increased waist along the x direction. Another waist can be reconstructed from ODT frequencies with the equation (4):

$$\frac{-4U_0}{m_{th}(2\pi)^2} = v_x^2 w_x^2 = v_z^2 w_z^2 = 2v_y^2 w_y^2 \qquad (6)$$

The waist of the sweeping ODT along the x direction is shown in Figure 3A. The sweeping shape was designed to make the beam profile parabolic [13] near its maximum. The position of the camera was scanned around the location of the focal spot to find the minimal beam size. Due to the large size of the swept beam, the laser spot occupied a large number of pixels. Its size was mostly determined by the sweeping amplitude rather than the quality of beam focusing, thus we excluded possible aberrations on the vacuum window and the additional mirror. The fit of the intensity profile measured this way returned a value for the beam width in the x-direction of $w_{x\ sweeping\ mod\ on} = 170 \pm 2\ \mu m$. Two standard deviations of the 1-D brightness profile of the swept beam (Figure 3A) gave a value of $189\ \mu m$ for the waist. (Standard deviation is estimated as $\sqrt{\sum_k (k-m)^2 I_k / \sum_k I_k}$ with a center of brightness position $m = \sum_k k \bullet I_k / \sum_k I_k$ with $I_k$ being the brightness of the bin number $k$.) Thus we estimated error, related to the difference between the observed swept beam profile and the Gaussian profile to be $19\ \mu m$. Finally, $w_{x\ sweeping\ mod\ on} = 170 \pm 19\ \mu m$.

Given this linear dimension, the rest of the measurements can be done via measurements of trap frequencies. The frequencies of the ODT were measured by the method described in the Frequency measurements chapter. The frequency measurements were done in 2 configurations: one with sweeping in the $x$ beam direction, and one without. This set of measures lead to following results:

$$\nu_{x\text{ sweeping mod on}} = 157 \pm 3 \text{ Hz}$$
$$\nu_{z\text{ sweeping mod on}} = 1649 \pm 6 \text{ Hz}$$
$$\nu_{x\text{ sweeping mod off}} = 981 \pm 6 \text{ Hz} \quad (7)$$
$$\nu_{z\text{ sweeping mod off}} = 1560 \pm 8 \text{ Hz}$$

Here error bars are statistical errors. Using equation (6) the parameters of the beam were found to be:

$$w_z = w_{x\text{ sweeping mod on}} \cdot \frac{\nu_{x\text{ sweeping mod on}}}{\nu_{z\text{ sweeping mod on}}} = 25.7 \pm 3.5 \, \mu m$$

$$w_x = w_z \cdot \frac{\nu_{z\text{ sweeping mod off}}}{\nu_{x\text{ sweeping mod off}}} = 15.8 \pm 2.3 \, \mu m \quad (8)$$

Finally, we measured the dependences of the frequencies in the two orthogonal directions versus power without sweeping regime (Figure 2D). The ratio between the found parameters $w_x$ and $w_z$ is constant with power much like it was expected: $w_z/w_x = 1.63 \pm 0.01$.

## C. Tensor and scalar polarizability

As it can be seen from equation (1), when $\theta_k = 90°$ or when the ellipticity parameter $\varepsilon = |\vec{u}^* \times \vec{u}| = 0$ (ODT-beam linear polarized) then the vector term in (1) becomes zero and therefore:

$$\alpha_{tot} = \alpha_{sc} + \frac{3m_J^2 - J(J+1)}{J(2J-1)} \times \frac{3\cos^2\theta_p - 1}{2} \alpha_t \quad (9)$$

The angle, $\theta_k$, was varied with adiabatic change of an external magnetic field governing atom orientation. Orientation of the magnetic field was calibrated with microwave spectroscopy [16]. In addition, the ODT-beam polarization was cleaned by a polarization beam splitter in combination with a $\lambda/4$ plate and thus the ellipticity parameter, $\varepsilon = |\vec{u}^* \times \vec{u}|$, was less than 0.03 during all the measurements.

To measure the tensor part of the polarizability, we used a linear polarized ODT varying angle of $\theta_p$ (see Figure 3B). Orientation of the ODT polarization was controlled by a half-lambda plate. For each position of the $\lambda/2$ plate, the ODT was loaded as described above and frequency of ODT in the $z$-direction was measured (see Frequency measurements). Polarization of the beam was checked by polarization beam splitter (PBS) placed after the vacuum chamber. The polarizability was calculated from measured frequencies using equation (5) and (9): $\alpha_{sc} = 547 \pm 51$ a.u. $\alpha_t = -145 \pm 14$ a.u. . While there is remarkable agreement between experimental and calculated values of tensor polarizability, the scalar part is slightly lower then the calculated value. We dedicate this deviation to systematic errors in the determination of the beam waist and its anharmonicity.

Besides measurements of polarizability, the lifetime of thulium atoms in the ODT was measured versus $\theta_p$ (see Figure 3C). The lifetime was extracted from decay of the number of atoms in ODT by fitting the curve with exponential dependence. One could see that lifetime mostly follows the real part of the polarizability. Here we see that lifetime is heavily affected by the light polarization being in antiphase with the photon scattering rate, which originated from the imaginry part of polarizability.

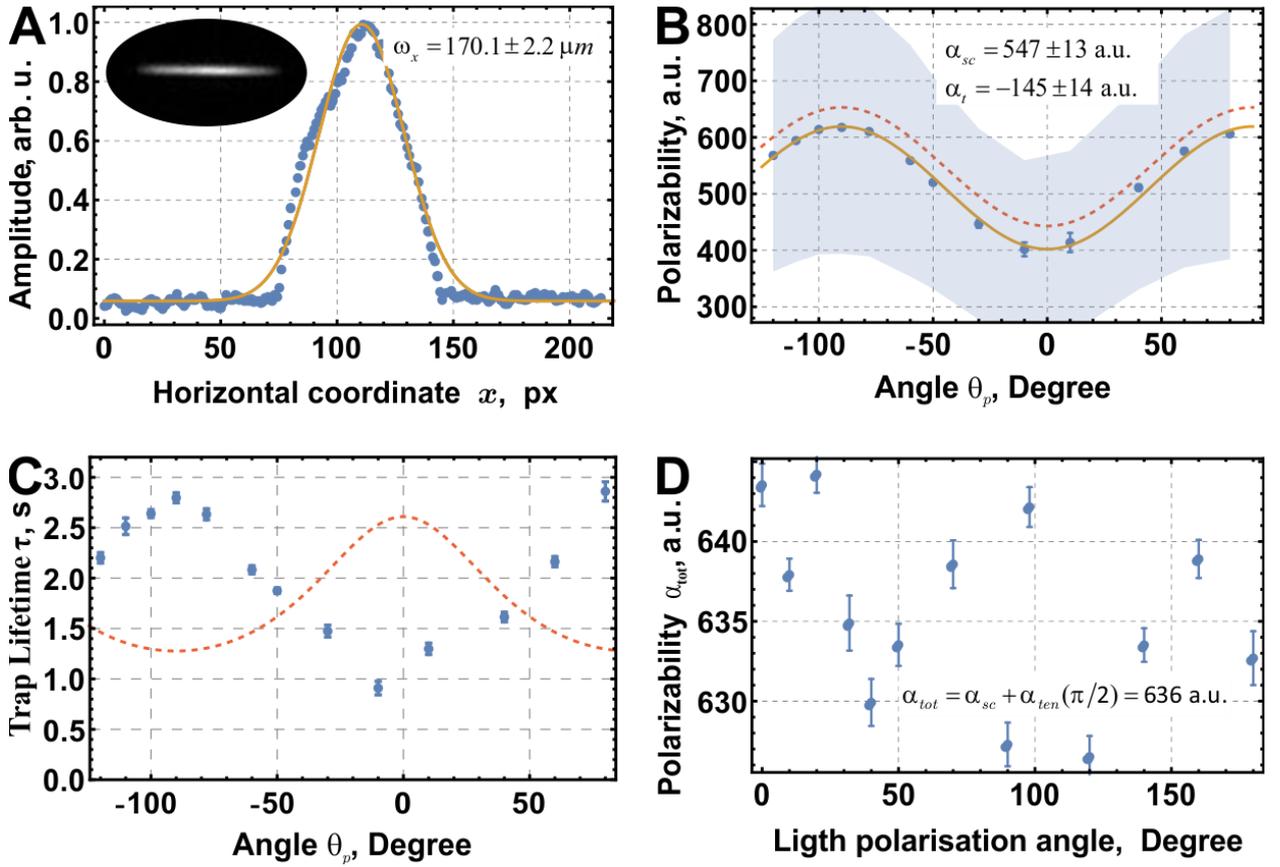

Figure 3 A – Sweeping trap profile (beam intensity was time averaged during imaging), dots represent intensity averaged by y dimension, solid yellow line represents gauss fit. B – Atomic Polarizability versus $\theta_p$. Gray area illustrates systematic uncertainty of the measurements. Solid yellow line represents fit of the experimental data with sine dependence, dashed Red line represents the simulations of the real part of the polarizability. C – Lifetime of atoms in the dipole trap versus $\theta_p$. Dashed red line represents normalized and inversed calculated imaginary part of the polarizability. D – Atomic polarizability versus the linear light polarization angle. In this experiment the magnetic field was codirected with the ODT beam and $\theta_k = 0°$.

Finally, to check how precisely the magnetic field is controlled, we set the magnetic field along the direction of light propagation. In this experiment, the magnetic field was codirected with the ODT beam and $\theta_k = 0°$, $\theta_p = 90°$; Thus, rotation of the light polarization does not change polarizability at all. The ODT linear light polarization was rotated using a $\lambda/2$ - plate. As one can see from Figure 3D, indeed the measured polarization does not change when light polarization changes and it is equal to the expected value from equation (9).

### D. Vector polarizability

As described above, with magnetic field aligned along beam propagation direction $\theta_k = 0°$, tensor polarizability does not depend on light polarization anymore. Therefore, this configuration is perfect for measurements of the vector part of polarizability. Thus, to determine vector polarizability, the magnetic field (3.94 G at this experiment) was aligned along the beam propagation direction and the ODT light ellipticity parameter, $\varepsilon = |\vec{u}^* \times \vec{u}|$, as well as sign of light polarization (sign of $\cos[\theta_k]$) was varied. This was realized by rotation of the $\lambda/4$ plate placed into the ODT beam. The ellipticity parameter was measured at all points by PBS and was calculated as:

$$\varepsilon = 2\frac{p}{1+p^2}$$
$$p = \sqrt{\frac{MinPower_1}{MaxPower_2}}, \quad (10)$$

where $MinPower_1$ and $MaxPower_2$ are the minimum and maximum of powers of the beam at 2 orthogonal orientations of PBS. The circular polarization was formed by the $\lambda/4$ plate; its sign was determined from an angle between the linear polarization of incoming light and the plate's fast axis.

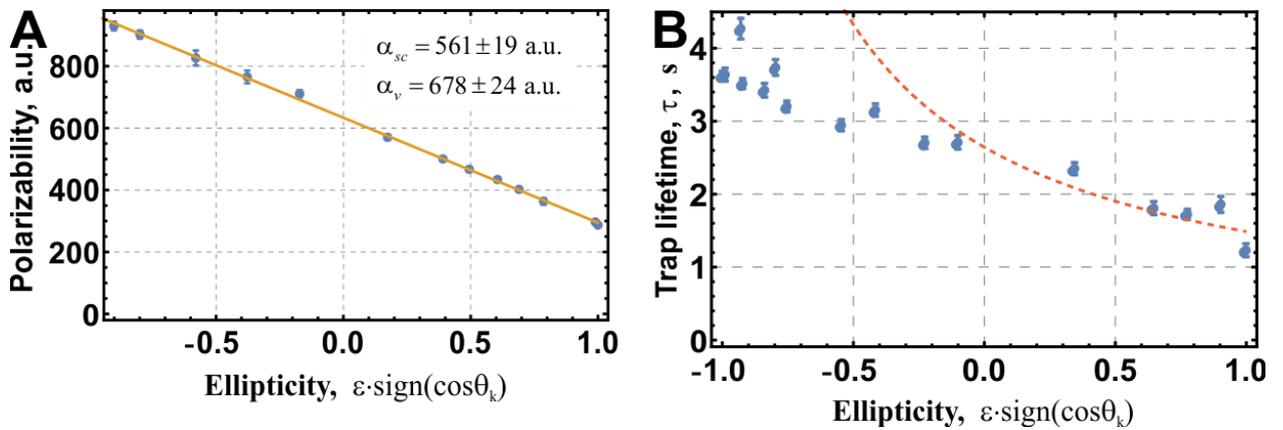

*Figure 4 A – Dependence of the atomic polarizability versus the ellipticity parameter and sing of circular polarization. At this experiment $\theta_k \approx 0°$. The plot shows statistical uncertainty only. B – Lifetime of the ODT versus ellipticity of the trap beam. Red dashed line indicates the normalized, inversed imaginary part of the polarizability.*

As is shown in Figure 4A, using fitted data from (1) with $\theta_p = 90°$, $\alpha_t = -145$ a.u., and keeping $\alpha_{sc}$ as a parameter produces almost the same scalar part of polarizability at $\alpha_{sc} = 561 \pm 19$ a.u. and $\alpha_v = 678 \pm 24$ a.u..

In order to relate the losses in the dipole trap to polarization, we also measured the dependence of the trap lifetime on the ellipticity parameter (see Figure 4B). It is interesting to note that if the vector part of polarizability again deviates from the behavior of the imaginary part and rather follows the real part; then contrary to the tensor case, the general trend is the same.

*Table 1. Polarizability of Tm atom near 532 nm.*

| Polarizability, Real part | Simulated polarizability, $\alpha^{theor}$ (a.u.) | Measured polarizability $\alpha^{expt}$ (a.u.) | Statistical uncertainty $\Delta\alpha_{stat}$ (a.u.) | Systematic uncertainty $\Delta\alpha_{sys}$ (a.u.) | Total uncertainty $\Delta\alpha_{tot}$ (a.u.) |
|---|---|---|---|---|---|
| Scalar | 583 | 547 | 13 | 190 | 190 |
| Tensor | -140 | -145 | 14 | 51 | 53 |
| Vector | 684 | 676 | 24 | 240 | 240 |

The values of the atomic polarizability of the thulium atom, including systematics uncertainties, (see APPENDIX B) are summarized in Table 1.

## V. CONCLUSION

Vector, tensor, and scalar polarizabilities were measured in thulium atoms ground state near 532.07 nm wavelength; which is a particularly important wavelength for optical dipole traps with Thulium atoms. Experimental values were compared to the theoretically calculated values and these were in nice agreement with each other. It was found that at this wavelength contributions of tensor and vector polarizability was quite significant, thus allowed for the formation of well-separated, spin-dependent lattices. Besides this, the losses of the optical dipole trap had also been measured, and these losses demonstrated correlation with the real rather than the imaginary parts of polarizability.

## VI. AKNOWLEDGEMENTS

This research was supported by the Russian Science Foundation grant #18-12-00266. We also thank Aubrey Sergeant for help with manuscript preparation.

## APPENDIX A: POLARAZABILITY

Full atomic polarizability was calculated by summing the contributions of all known polarizabilities from the National Institute of Standards' database [11] and using formulas (4,5,6) from [17]. In the arbitrary case, angular dependence of tensor part of polarizability is different from expression (1) and is given by [18,19]:

$$f(\theta_k, \theta_p, \varepsilon) = 1 - (3/2)sin(\theta_k)^2(1 + \sqrt{1-\varepsilon^2}cos(2\theta_p)), \quad (11)$$

where $f(\theta_k, \theta_p, \varepsilon)$ is function, determining the angular dependence. That leads to $f(\theta_p) = 1 - 3cos(\theta_p)^2$ in the case of linearly polarized light with a wave vector perpendicular to the magnetic field direction ($\theta_k = \pi/2$), and degenerates to $f = 1$ in the case of light propagating in $z$ direction ($\theta_k = 0$).

## APPENDIX B: SYSTEMATIC UNSERTAINTY

The systematic uncertainty was estimated as following:

$$\frac{\Delta \alpha_{sys}}{\alpha} = \sqrt{\left(\frac{\Delta w_{z\,sys}}{w_z}\right)^2 + \left(3\frac{\Delta w_{x\,sys}}{w_x}\right)^2 + \left(\frac{\Delta P_{sys}}{P}\right)^2}.$$

We estimate a power uncertainty of 2%. However, the uncertainty in measuring waists (about 11%) made a major contribution to systematic uncertainty. Summing up all sources of uncertainty, the final systematic uncertainty for the measured scalar, tensor and vector polarization is about 35%.